\documentclass[letterpaper]{article}

\usepackage{natbib,alifeconf}  

\title{Criticality as It Could Be: organizational invariance as self-organized criticality in embodied agents}
\author{Miguel Aguilera$^{1,2,3}$ \and Manuel G. Bedia$^{1,2}$ \\
\mbox{}\\
$^1$Dept. of Computer Science, Univ. of Zaragoza, Zaragoza (Spain) \\
$^2$Arag\'on Institute of Engineering Research, Zaragoza (Spain) \\
$^2$Institute for Cross-Disciplinary Physics and Complex Systems, Palma (Spain)\\
sci@maguilera.net} 

\usepackage{mathtools}
\usepackage{hyperref}

\begin{document}
\maketitle

\begin{abstract}

This paper outlines a methodological approach for designing adaptive agents driving themselves near points of criticality. Using a synthetic approach we construct a conceptual model that, instead of specifying mechanistic requirements to generate criticality, exploits the maintenance of an organizational structure capable of reproducing critical behavior. 
Our approach exploits the well-known principle of universality, which classifies critical phenomena inside a few universality classes of systems independently of their specific mechanisms or topologies. 
In particular, we implement an artificial embodied agent controlled by a neural network maintaining a correlation structure randomly sampled from a lattice Ising model at a critical point. 
We evaluate the agent in two classical reinforcement learning scenarios: the Mountain Car benchmark and the Acrobot double pendulum, finding that in both cases the neural controller reaches a point of criticality, which coincides with a transition point between two regimes of the agent's behaviour, maximizing the mutual information between neurons and sensorimotor patterns. 
Finally, we discuss the possible applications of this synthetic approach to the comprehension of deeper principles connected to the pervasive presence of criticality in biological and cognitive systems. 

\end{abstract}

\section{Introduction}

Biological systems at a wide range of scales -- from protein families to brains -- show signatures of criticality. These systems do not operate deep into one or other regime of activity instead of they operate at or near critical points, poised at transitions in their parameter space \citep{mora_are_2011}. In a nutshell, criticality refers to a distinctive set of properties found at the boundary separating regimes with different dynamics: the transition between an ordered and a disordered phase. Some of these properties include power-law divergences of some quantities described by critical exponents and fractal behaviour \citep{salinas_scaling_2001}.
Signatures of criticality have been detected in neural cultures \citep{schneidman_weak_2006}, immune receptor proteins \citep{mora_maximum_2010}, the network of genes controlling morphogenesis in fly embryos \citep{krotov_morphogenesis_2014} or lipid membranes \citep{honerkamp-smith_introduction_2009}. As well, indicators of critical behaviour have been observed in the human brain \citep{chialvo_critical_2014} and cognitive behavioural patterns \citep{van_orden_blue-collar_2012}. These results hint at general theoretical principles underlying biological self-organization, compelling us to ask what type of mechanisms are driving biological systems at a dauntingly diverse range of levels of organization to operate near critical points of activity.

This question is largely unresolved. 
Firstly, because the connection between experimental descriptive indicators of criticality and models is often fragile \citep{wagenmakers_abstract_2012}. 
On the other hand, models of criticality are generally used at the level of analogy, based on specific  mechanistic requirements reproducing criticality in a set of particular cases, providing a myriad of different models but often failing to capture explanations of a few more universal classes and properties. During the last decade a new generation of experiments has attempted to go beyond analogies generating models directly inferred from biological data \citep{mora_are_2011}. However, the difficulty of inferring general principles from the specific modelled mechanisms is still patent in these models.


In order to progress in the comprehension of critical phenomena, there are some alternatives and suggestions that we could explore with more emphasis. For example, we could try to detect universal mechanisms able to generate critical activity in a wide set of systems and contexts. This approach -- a kind of `Artificial Life route' to self-organized criticality -- could promote the development of conceptual models explaining how organisms are driven towards criticality in an abstract level, working as `proofs of concept' \citep{barandiaran_animats_2009} and supporting existing and future experimental findings.
Specifically, here we focus on the relation of criticality with organizational invariants of the model. 
By this, we mean that instead of specifying specific mechanistic requirements to generate criticality, we explore the possibility of generating criticality through the conservation of certain invariants in the relations between the components of the system (e.g. correlations between components), i.e. imposing certain patterns in the system's organization.
This organizational point of view, in contrast with focusing on  intrinsic properties of the components of the system, is supported by the existence of well-known universality classes that provide a unified expression for families of systems operating under criticality \citep{kadanoff_more_2009}. Systems under the same universality class, even if they are defined by very different material parameters or physical properties, present the same critical exponents characterizing diverging observables which are defined by the symmetries of the system.

In line with these ideas, the paper is structured as follows. First, we propose a conceptual model based on statistical mechanics to design an artificial agent that maintains organizational invariants in its structure, specifically a distribution of correlations between the components of the system. Operatively, the model is implemented as a `Boltzmann machine' neural network reproducing a correlation structure randomly sampled from a 2D lattice Ising model at a critical point. 
Then, we test the model in two classical examples of learning and control: the Mountain Car and the double pendulum. In both examples we find that agents with no free parameters exploit the whole dynamic range of available configurations being poised near critical transition points of their parameter space between qualitatively different behavioural regimes. Finally, we discuss the possible applications of this synthetic approach to contribute to the comprehension of the deeper principles that governs biological and cognitive systems.

\section{Organizational invariants of self-organized criticality}

The development of mechanistic models of criticality in biological systems -- based on finding the particular properties and generative mechanism that give support this regime of behaviour -- typically assumes a particular way of modelling. Typically, we can find models of criticality exploiting clever local rules generating critical behaviour \citep[e.g.][]{bak_self-organized_1987, bak_punctuated_1993}, parameter tuning of systems showing critical phase transitions \citep[e.g.][]{beggs_neuronal_2003, kitzbichler_broadband_2009} and/or the use of specific structural properties of the underlying network of the system \citep[e.g.][]{rubinov_neurobiologically_2011, haimovici_brain_2013}. However, as we have pointed out above, families of critical phenomena can be classified into universality classes determined only by a few properties of the system. 
One of these properties is that all the critical exponents of all models within a given universality class are exactly the same. For example, in all Ising models in 2D lattices (square, triangular, hexagonal and so forth) spin-spin correlations follow the asymptotic form $c(r) \propto 1/r^{\eta}$, where $\eta=1/4$ \citep{salinas_scaling_2001}. This surprising property provides a perspective about criticality in terms of universal relations, suggesting that we could model criticality using simple and non-specific mechanisms independently on the topology of the system. 

Following this intuition, we propose a model not committed to a particular local behavioural rule or connectivity topologies. Instead, our model is built on the maintenance of an organizational structure for reproducing the global functional properties of a family of critical models.  There exists some experimental evidence showing that, given a Ising model near a critical point, one could build a family of models by learning correlations drawn at random from the original model, which will be poised near a similar critical point \citep{tkacik_spin_2009}. Based on this inspiration, we propose to reproduce and support criticality by the invariant maintenance of an organizational structure of correlations following a $1/r^\eta$ law. Thus, instead of restricting a model to a particular set of mechanisms, we find a general organizational distribution which could be easily implemented by very different structures, driving a system to a regime of criticality.

We define our model as a neural network as an Ising model or Boltzmann Machine \citep{ackley_learning_1985} following a maximum entropy distribution:
\begin{equation}
 P(s) =   \frac{1}{Z} exp \Bigg[ {\beta \sum_i h_i s_i + \sum_{i < j} J_{ij} s_i s_j}\Bigg]
 \label{eq:Ising}
\end{equation}
where the distribution follows a en exponential family $P(s) = \frac{1}{Z} e^{-\beta E(s)}$, $Z$ is a normalization value,
where the energy $E(s)$ of each state is defined in terms of the bias  $h_i$  and couplings  $J_{ij}$  between pairs of units, and  $\beta = 1/({T k_B})$, being $k_B$ Boltzmann's constant and $T$ the temperature of the system. 
Units $s_i$ can take discrete values of $+1$ or $-1$ and the couplings and bias can take continuous values.
Without loss of generality we can set an operating working temperature such that $\beta= 1$.
To simulate the network, we use Glauber dynamics, by which the value of a unit is inverted each time with probability:

\begin{equation}
 P_i(s) = \Bigg[ 1 + e^{\beta \Delta E_i(s)}\Bigg]^{-1}
\end{equation}

where $\Delta E_i(s) = 2(h_i s_i + \sum_j J_{ij} s_i s_j)$ is the energy difference between the inverted and original state. Throughout the paper, we simulate the network updating its state sequentially by applying Glauber dynamics to all units in the network in a random order at each simulation step.

\begin{figure}[ht]
\begin{center}
 \begin{tabular}{c}
   \multicolumn{1}{l}{\textbf{A}} \\
  \includegraphics[width=5.0cm]{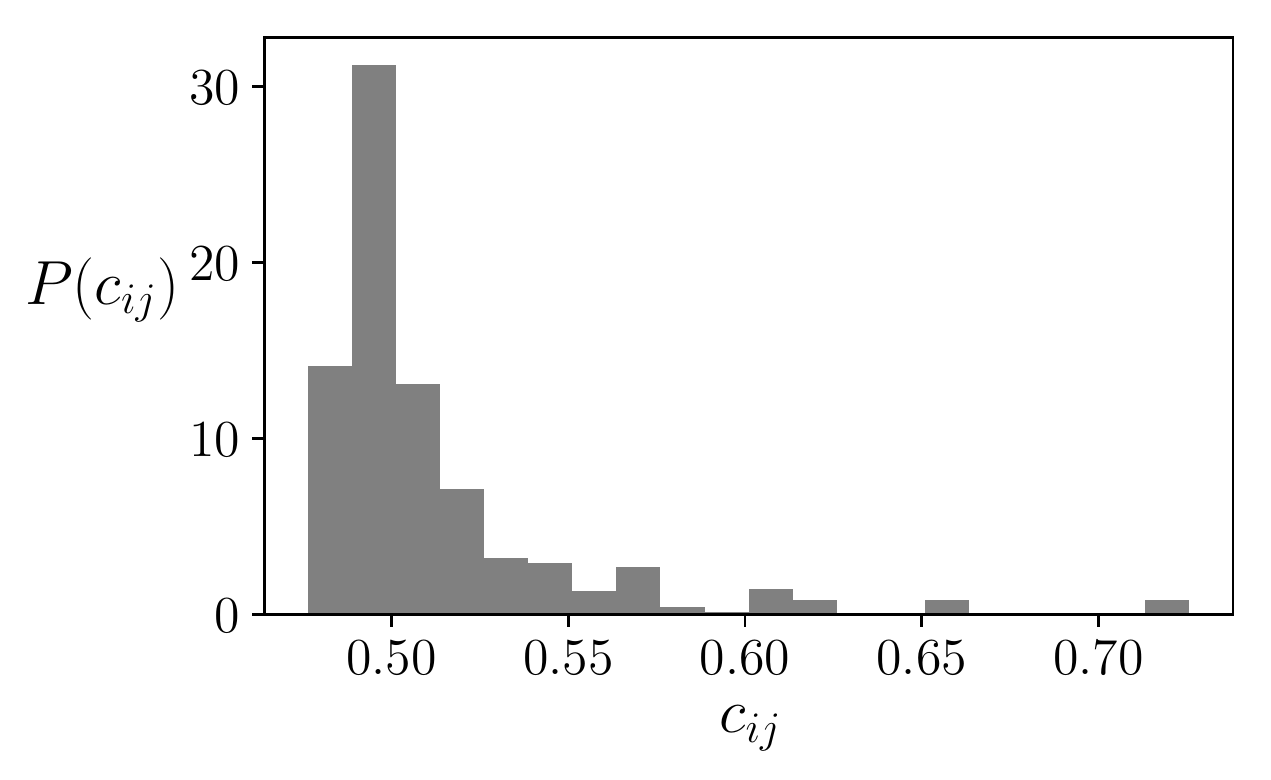} \\
  \multicolumn{1}{l}{\textbf{B}}  \\
  \includegraphics[width=5.0cm]{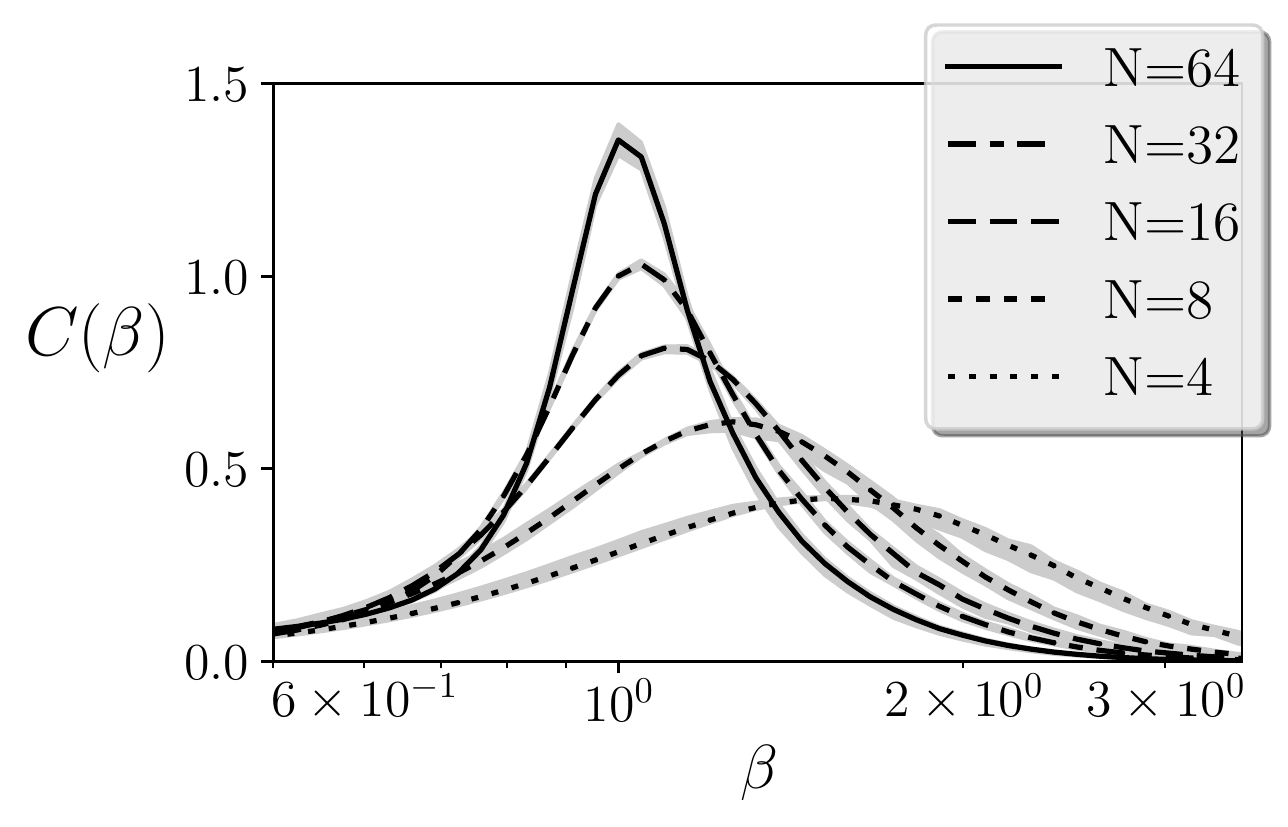}
 \end{tabular}
 \begin{tabular}{ll}
     \multicolumn{1}{l}{\textbf{C}}  &  \multicolumn{1}{l}{\textbf{D}} \\
  \includegraphics[width=3.5cm]{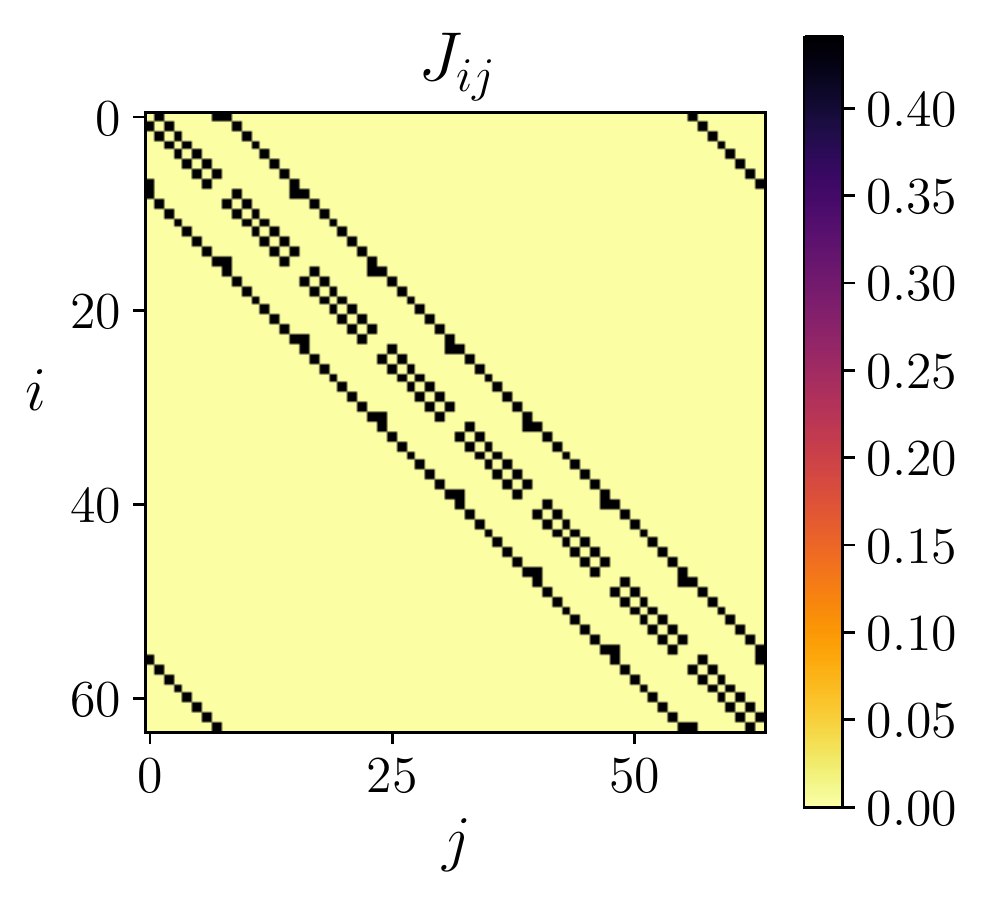} &  \includegraphics[width=3.5cm]{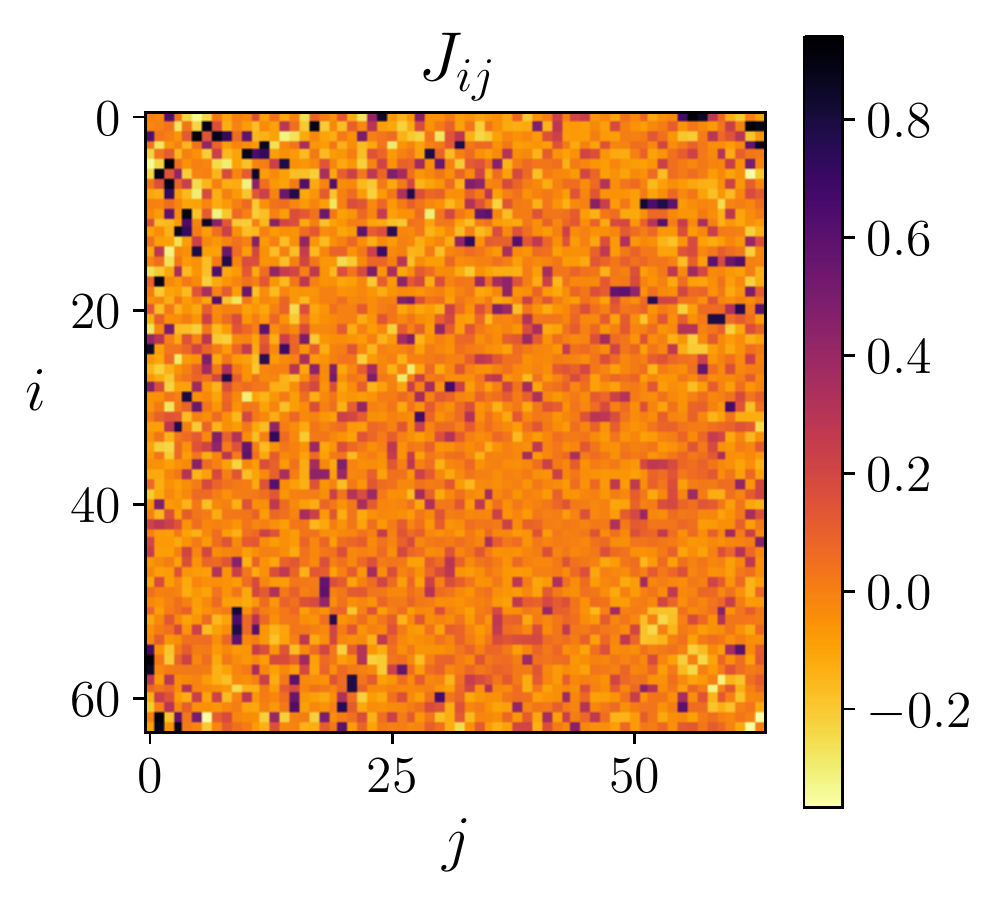}
 \end{tabular}
\end{center}
\caption{\textbf{(A)} Distribution of correlation values used for learning, extracted from a 20x20 lattice Ising model at critical temperature. \textbf{(B)} Divergence of the heat capacity in 10 models learning random correlations sampled from Figure \ref{fig:Correlations}.A. Maximum and minimum values are shown by the grey area. \textbf{(C)} Distribution of the coupling matrix of a 8x8 lattice Ising model with periodic boundaries. \textbf{(D)} Distribution of a 64 units Ising model learning correlations sampled from Figure \ref{fig:Correlations}.A. The order of the nodes in the coupling matrix has been rearranged using hierarchical clustering.} 
\label{fig:Correlations}
\end{figure}
 
In order to induce criticality in its behaviour, our model maintains a specific internal structure of correlations, conserving the statistical properties of a system that we know it is at criticality. 
Since the size of our models will be far from the thermodynamic limit, instead of directly using the asymptotic form $c(r) \propto 1/r^{\eta}$, we approximate it by computing the correlation structure of a known model showing this distribution in the thermodynamic limit. 
One of the few specific cases where the Ising model presents an exact solution is a model with zero fields and 2D lattice connectivity, in which a critical point appears at $J=\beta log(1+\sqrt{2})/2$ \citep{onsager_crystal_1944}. 
Specifically, we use a 20x20 2D lattice Ising model at critical temperature with periodic boundary conditions. 
We simulate the model using Glauber Dynamics, generating $10^6$ samples, after initializing the model running $10^5$ updates. From this simulation, we obtain the distribution of correlations in the system $c_{ij}=\langle s_i s_j\rangle$ observed in Figure \ref{fig:Correlations}.A. Since the fields at all units are zero, the means $m_i = \langle s_i \rangle$ of all units are also zero.

Once obtained a distribution of correlations, we will generate new models by assigning to each unit and pair of units means and correlations randomly drawn at random from the distribution of the 20x20 Ising model.
At this point, the problem is that it is not trivial finding which combination of $h_i$ and $J_{ij}$ generates a specific combination of $m_j$ and $c_{ij}$. This is known as the `inverse Ising problem'. This problem was formulated by \cite{ackley_learning_1985} in their discussion of Boltzmann machines and can be solved by a simple gradient descent rule:

\begin{equation}
 \begin{split}
  h_{i} \leftarrow h_{i} + \mu  (m_i - m_i^m )\\
  J_{ji} \leftarrow J_{ji} + \mu  ( c_{ij} - c_{ij}^m )     
 \end{split}
  \label{eq:learning}
\end{equation}
where $\mu$ is a constant learning rate, $m_{i}$ and $c_{ij}$ are the objective mean and correlations of the learning algorithm, and $m_{i}^m$ and $c_{ij}^m$ are the mean and correlations of the model for the current values of $h_i$ and $J_{ij}$. 
Computing each learn step is computationally costly, since it requires to sum over all possible states of $s$, although approximate methods as Monte Carlo sampling are generally used to speed up learning.
As a demonstration, we apply the learning rule to different models assigning them objective means and correlations drawn at random from the distribution found for the 20x20 lattice Ising model.
For each model, we apply Equation \ref{eq:learning} for learning $m_{i}$ and $c_{ij}$, computing the actual $m_{i}^m$ and $c_{ij}^m$ with Glauber dynamics.
Since precision of learning is not important (the objective is to capture the overall distribution) we do not wait for convergence of the algorithm and simply update the learning rule $1000$ times. We used a learning rate $\mu=0.01$ and computed $1000 N$ samples for each learning step, being $N$ the size of the system.

Using this method, we apply the learning rule to 10 different models for sizes $N=4, 8, 16, 32, 64$. For each model, we test if the models are at criticality by computing their heat capacity  $C(\beta) = \beta^2 \langle E^2(s) \rangle - \langle E(s) \rangle ^2$, where $E(s) = - \sum_i h_i s_i - \sum_{i < j} J_{ij} s_i s_j$ is the energy of the Ising model. The divergence of the heat capacity is an indicator of criticality. We simulate each model for $10^5$ steps for different values of $\beta$, and we found that all the 10 models diverge at the operating temperature of $\beta=1$ (Figure \ref{fig:Correlations}.B), showing similar values of heat capacity to the original lattice Ising model with periodic boundaries.
Nevertheless, although the distribution of correlations is similar than the lattice Ising model, the structure of the model is radically changed. Instead of the original ordered structure of a uniform lattice (Figure \ref{fig:Correlations}.C), we have now a disordered distribution of couplings $J_{ij}$ (Figure \ref{fig:Correlations}.D), including both positive and negative couplings. Also, each random selection of correlations yields a completely different arrangement of values of couplings $J_{ij}$.

In the following section, we use this learning rule  to drive the neural controller of an embodied agent towards a critical point. In order to do so, we need to take into account the environment during learning. If we consider two interconnected Ising models, (one being the neural controller and other being the environment) Equation \ref{eq:learning} holds perfectly if we only apply it to the values of $i$ and $j$ corresponding to units of the neural controller.
In our case, we will not use an Ising model as an environment but instead we will use two classical examples from reinforced learning. Therefore, our learning rule will be valid as long as the statistics of the environment can be approximated by an Ising model with an arbitrary number of units. Luckily, Ising models are universal approximators \citep{montufar_universal_2014} and any arbitrary environment could be approximated by an equivalent Ising model.

\begin{figure}[ht]
\begin{center}
 \begin{tabular}{c}
  \multicolumn{1}{l}{\textbf{A}}  \\
 \frame{\includegraphics[width=4.0cm]{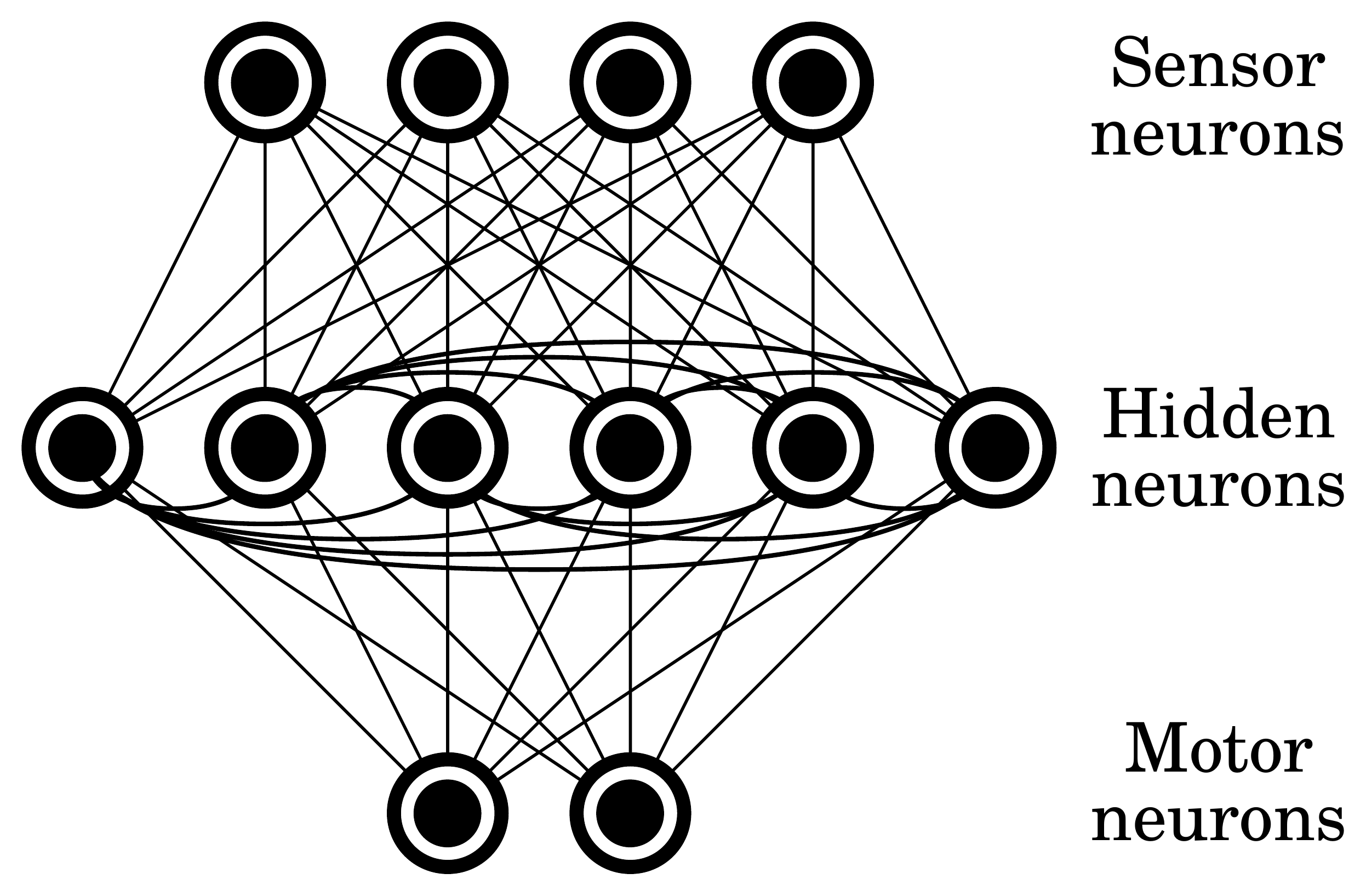}}  \\
  \multicolumn{1}{l}{\textbf{B}}  \\
  \frame{\includegraphics[width=4.0cm]{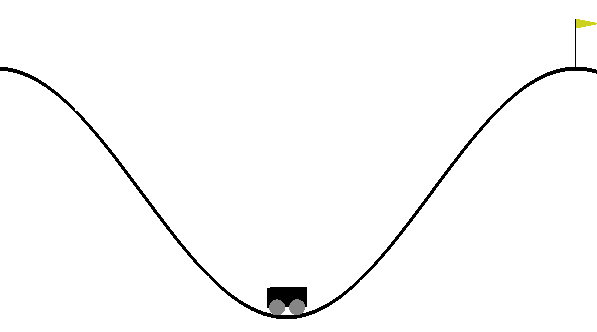}}
  \\
  \multicolumn{1}{l}{\textbf{C}} \\
  \frame{\includegraphics[width=4.0cm]{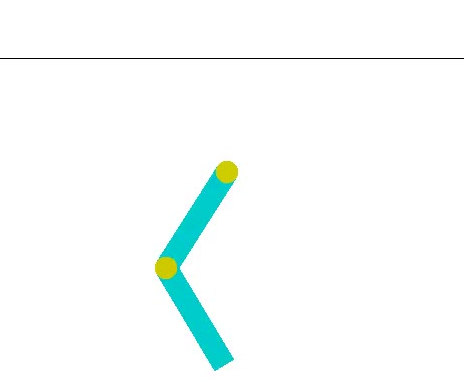}}
 \end{tabular}
 
\end{center}
\caption{(\textbf{A}) Structure of the embodied neural controller. (\textbf{B}) Mountain Car environment: an under-powered car that must drive up a steep hill by balancing itself to gain momentum. (\textbf{C}) Acrobot environment: an agent has to balance a double pendulum to reach the high part of the environment.} 
\label{fig:Embodiment}
\end{figure}

\section{Embodied model: Mountain Car and double pendulum}
In order to evaluate the behaviour of the proposed learning model, we test it in two embodied situations using the OpenAI Gym toolkit \citep{brockman_openai_2016}. 
For doing so we define a neural network consisting on an Ising model defined as in Equation \ref{eq:Ising}, describing a network of $8$ units, with $6$ hidden units and $2$ motor units, and a variable number of sensor units (4 or 6, depending on the environment).
Motor units will define the actions performed by the agents, while sensor units are updated directly from the state of the environment.
Sensor units and motor units are only connected to hidden neurons, while hidden neurons are fully connected to the rest of the system (Figure \ref{fig:Embodiment}.A). All connections are assigned an objective $c_{ij}$ value (selected at random from distribution $P(c_{ij})$ at Figure \ref{fig:Correlations}.A), and all units except sensor units are assigned an objective mean value of $m_i = 0$. During learning, the agent will apply the rule in Equation \ref{eq:learning} for adjusting its means and correlations to the objective values.
At each simulation step, the Ising model is simulated by updating its units in uniform random order using Glauber dynamics.
The first embodiment of the network consists in the Mountain Car environment \citep{moore_efficient_1990}. This environment is a classical testbed in reinforcement learning depicting an under-powered car that must drive up a steep hill (Figure \ref{fig:Embodiment}.B). Since gravity is stronger than the car's engine, the vehicle must learn to leverage potential energy by driving to the opposite hill before the car is able to make it to the goal at the top of the rightmost hill.
In this environment, the horizontal position $x$ of the car is limited to an interval of $[-1.5\pi,0.5\pi]$, and the vertical position of the car is defined as $y=\sin(3x)$. The velocity in the horizontal axis is updated each time step as $v(t+1)=v(t) + 0.001 a - 0.0025 \cos(3x)$, where $a$ is the action of the motor which can be either ${-1,0,1}$.
The sensors of the neural network are defined as an array of four units, which are fed the instantaneous velocity of the car, discretized into an array of 4 bits.  Each sensor unit is assigned a value of $1$ if its corresponding bit is active and $-1$ otherwise.

The second embodiment consists in a double pendulum or `Acrobot' \citep{sutton_generalization_1996} which has to coordinate the movements of two connected links to lift its weight (Figure \ref{fig:Embodiment}.C). The position of the system is defined by the angles of both pendulums $\theta_1$ and $\theta_2$, whose behaviour is defined by the following system of differential equations:
\begin{equation}
\begin{split}
\ddot \theta_{1} = -(d_{2} \ddot  \theta_{2} + \phi_{1}) / d_{1} \\
\ddot \theta_2=(m_2l_{c2}^2 + I_2 - \frac{d_2^2}{d_1})^{-1} (\tau + \frac{d_2}{d_1}\phi_1 - \phi_2)\\
d_{1} = m_{1} l_{c1}^2 + m_{2} (l_{1}^2 + l_{c2}^2 +\\+ 2 l_{1} l_{c2} \cos(\theta_{2})) + I_{1} + I_{2} \\
d_{2} = m_{2} (l_{c2}^2 + l_{1} l_{c2} \cos(theta_{2})) + I_{2} \\
\phi_{2} = m_{2} l_{c2} g \cos(\theta_{1} + \theta_{2} - \pi / 2) \\
\phi_{1} = - m_{2} l_{1} l_{c2} \dot \theta_{2}^2 \sin(\theta_{2}) -\\- 2 m_{2} l_{1} l_{c2} \dot \theta_{2} \dot \theta_{1} \sin(\theta_{2}) +\\
+ (m_{1} l_{c1} + m_{2} l_{1}) g \cos(\theta_{1} - \pi / 2) + \phi_{2}
\end{split}
\end{equation}
where $\tau$ is the torque applied to the system which can be either ${-1,0,1}$, $m_1=m_2=m$ is the mass of the links, $l_1=l_2=1$ is the length of the links and $l_{c1}=l_{c2}=0.5$ are the lengths to the center of mass of the links, and $I_1=I_2=1$ are the moments of inertia of the links and $g=9.8$ is the gravity.

The Acrobot embodiment was defined with 6 sensor units, divided in two groups of 3. Each group encoded the binarized horizontal and vertical positions of the tip of the second link, defined as $x=\sin(\theta_1) + \sin(\theta_1 + \theta_2), y=-\cos(\theta_1)-\cos(\theta_1+\theta_2)$.

In order to make the tasks challenging, we set the maximum velocity allowed to the Mountain car to $\pm 0.045$ (typically is set to $0.07$) and the mass of the Acrobot links to $m=1.75$ (typically $m=1$). These parameters are designed to make it difficult for agents controlled by neural networks with random parameters solve the task (reaching the top of the environment), having success rates of  $11.2\%$  for the Mountain Car and $7.7\%$ for the 
Acrobot\footnote{Success rates were evaluated by simulating 1000 neural controllers with random parameters (sampled from a uniform distribution in the range $[-2,2]$). The Mountain Car was simulated for 1000 simulation steps starting from a random position in the valley between $[0.4,0.6]$, and was considered successful when reached the maximum position at least once. The Acrobot was simulated for 5000 simulation steps from the bottom position (angles and angular speeds between $[-0.1,01]$) and was consider successful if reached a vertical position higher than $1.8$.}.

We train 10 agents for each embodiment applying the learning rule from Equation \ref{eq:learning}, with $\eta=0.01$. In order to avoid overfitting, we add an $L2$ regularization term $\lambda=0.004$. 
Note that agents during learning have no explicit goal other than adjusting the correlations of the system to a random sample extracted from the probability distribution in Figure \ref{fig:Correlations}.A.
Agents are initialized in the starting random positions in the bottom of their environments ($x\in [0.4,0.6]$ for the Mountain Car and $\theta_1,\theta_2,\dot\theta_1,\dot\theta_2  \in [-0.1,01]$ for the Acrobot). The state of the neural network is randomized and the initial parameters $h_i$ and $J_{ij}$ are set to zero. 
Agents are simulated for $1000$ trials of $5000$ steps, computing each trial the values of $m_{i}^m$ and $c_{ij}^m$ and applying Equation \ref{eq:learning} at the end of the trial . 
Note that agents are not reset at the end of the trial.

\section{Results}
In this section, we analyze the behaviour of the neural controllers and the behavioural patterns of the agents respect to the possibilities of their parameter space. 
The first striking result is that all 10 agents present quite similar behaviour for each embodiment, despite the fact that each one has learned different values of $c_{ij}$ and $J_{ij}$.
In order to compare the agents with other behavioural possibilities, we explore the parameter space by changing the parameter $\beta$ of the agents. Modifying the value of $\beta$ is equivalent to a global rescaling of the parameters of the agent transforming $h_{i} \leftarrow  \beta \cdot h_{i}$ and $J_{ij} \leftarrow \beta \cdot J_{ij}$, thus exploring the parameter space along one specific direction. For $21$ values of $\beta$ logarithmically distributed in the interval $[10^{-1},10^1]$ we simulate the 10 agents for each embodiment during $10^6$ simulation steps, after starting the agents from the random starting position and an initial run of $10^4$ simulation steps. We will use the results of those simulations for all the situations in this section.

\subsection{Signatures of criticality in the neural controller}
\begin{figure}[ht]
\begin{center}
 \begin{tabular}{ll}
  \textbf{A} & \textbf{B} \\
 \includegraphics[width=3.8cm]{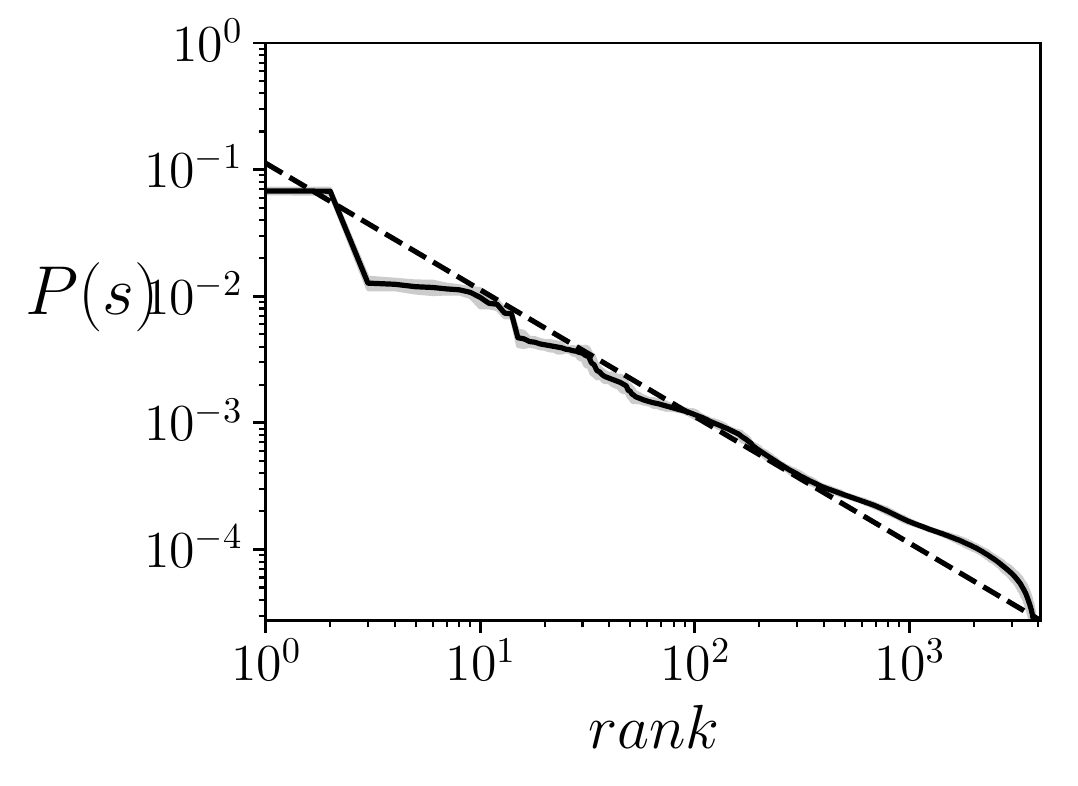} &
 \includegraphics[width=3.8cm]{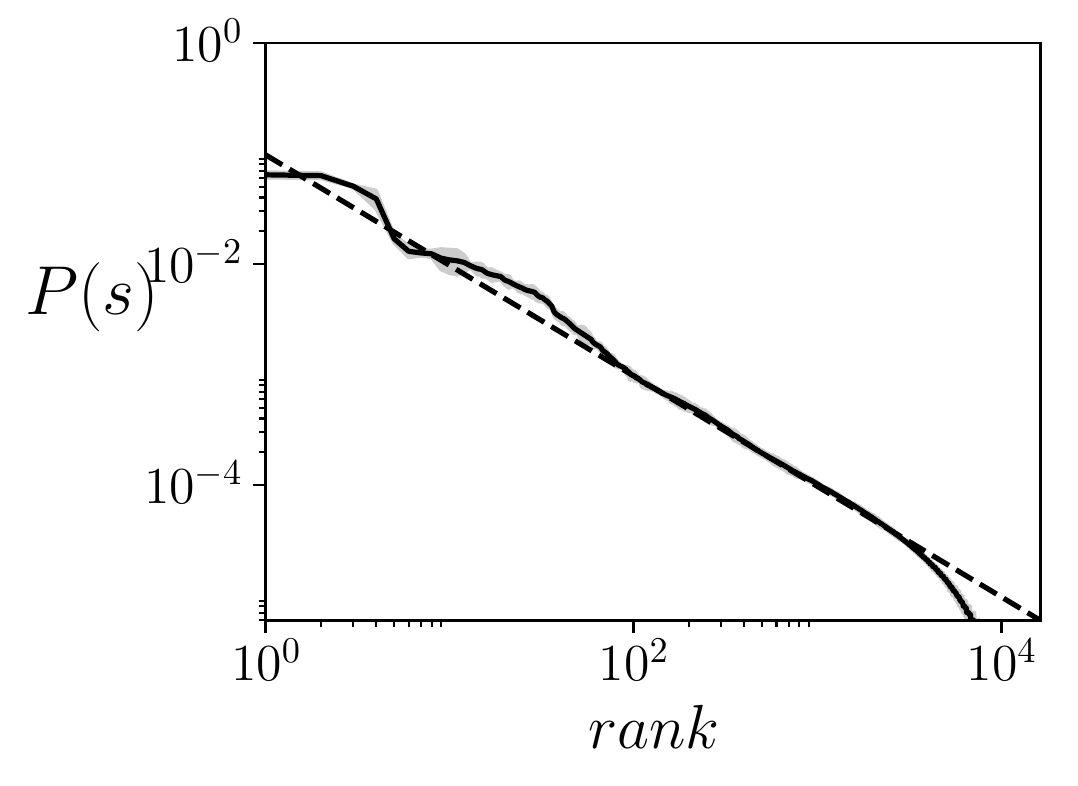} \\
  \textbf{C} & \textbf{D} \\
 \includegraphics[width=3.8cm]{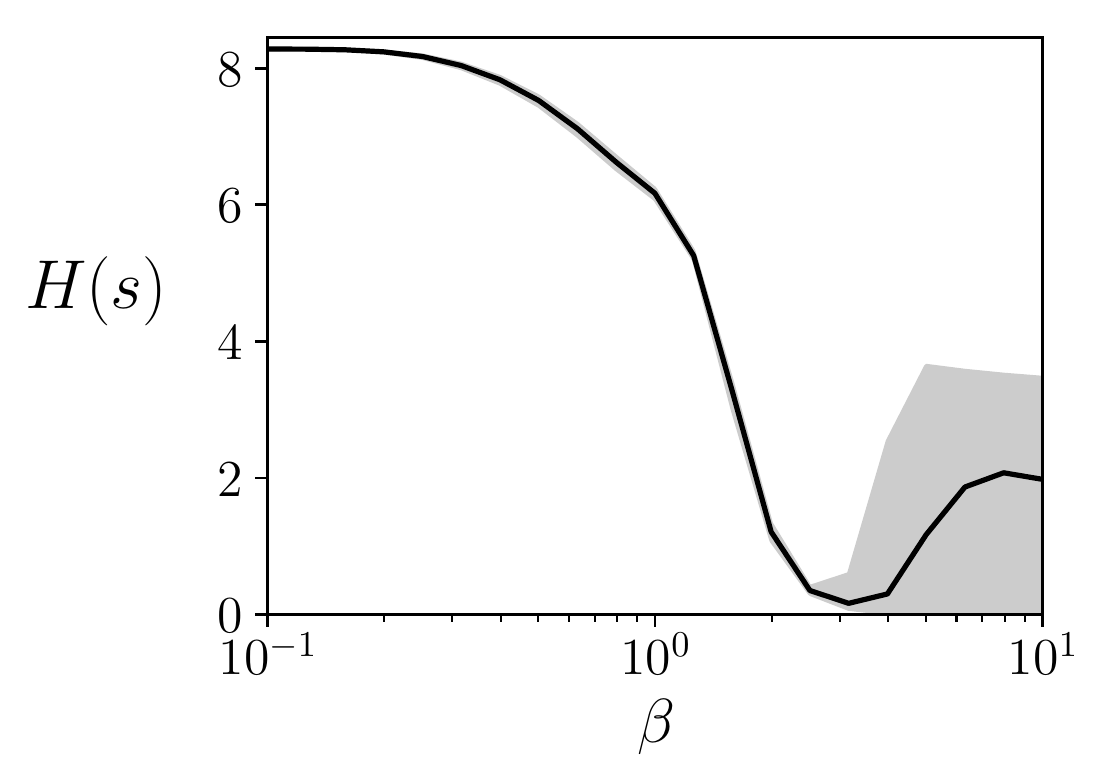} &
 \includegraphics[width=3.8cm]{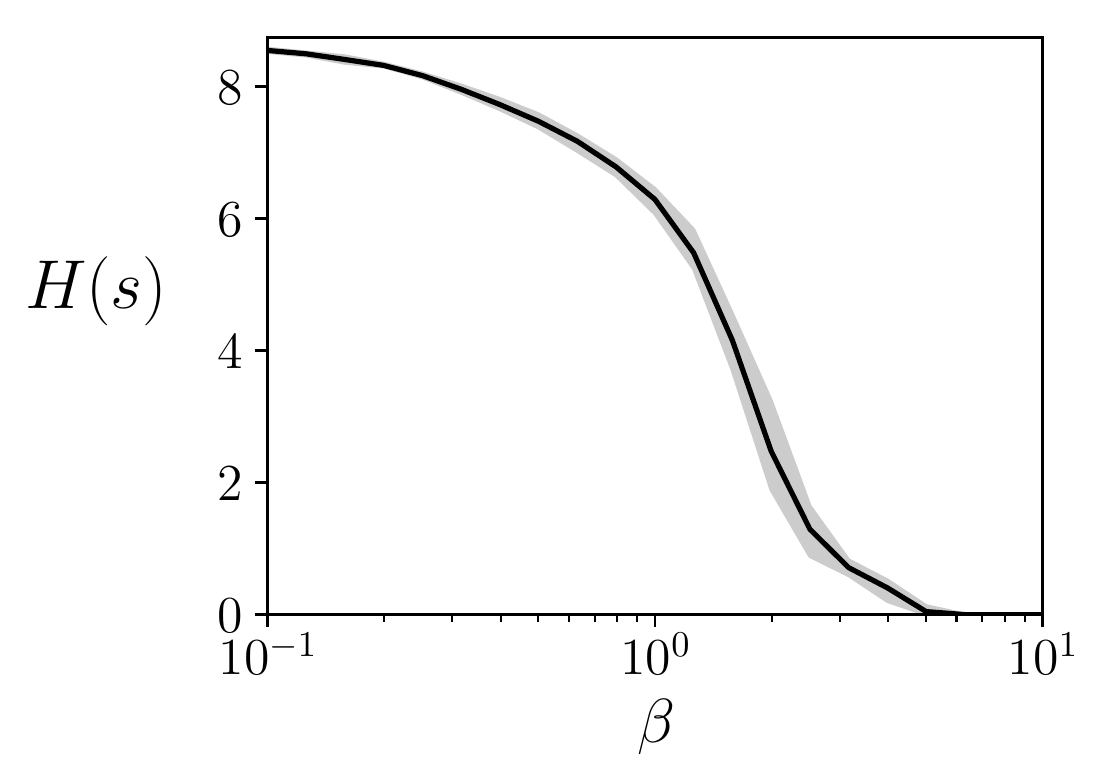} \\
   \textbf{E} & \textbf{F} \\
 \includegraphics[width=3.8cm]{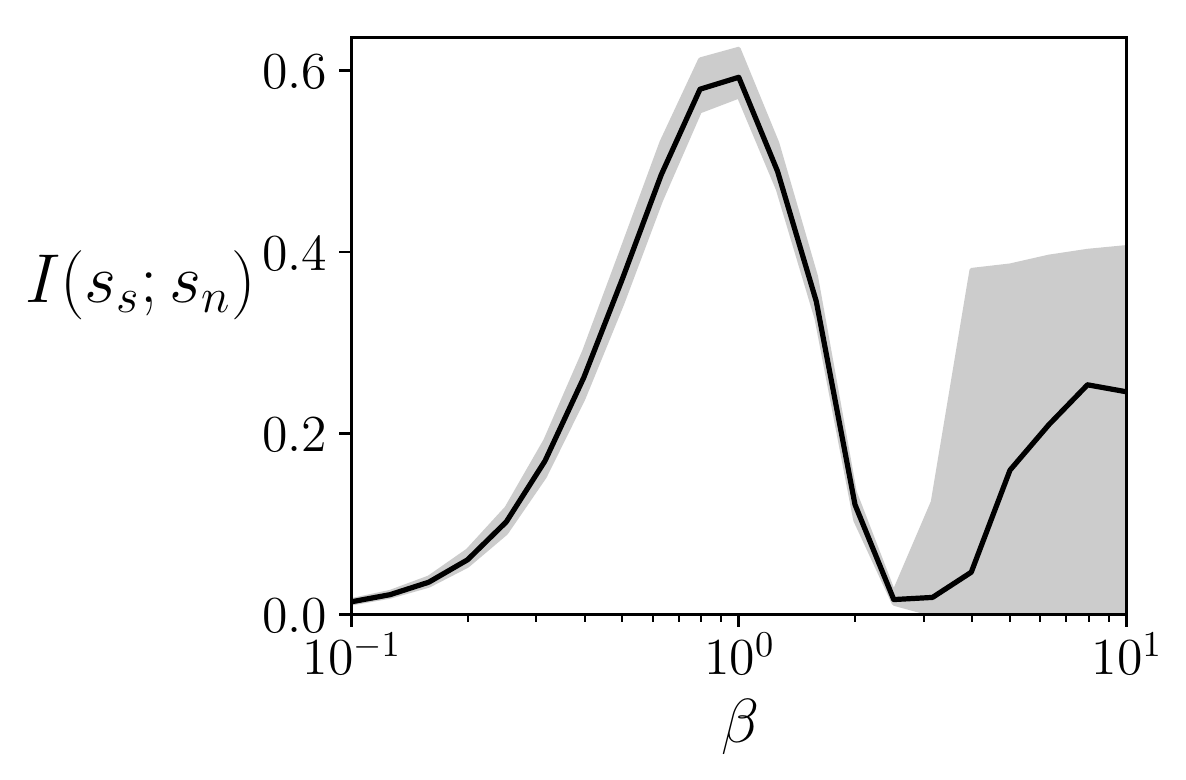} &
 \includegraphics[width=3.8cm]{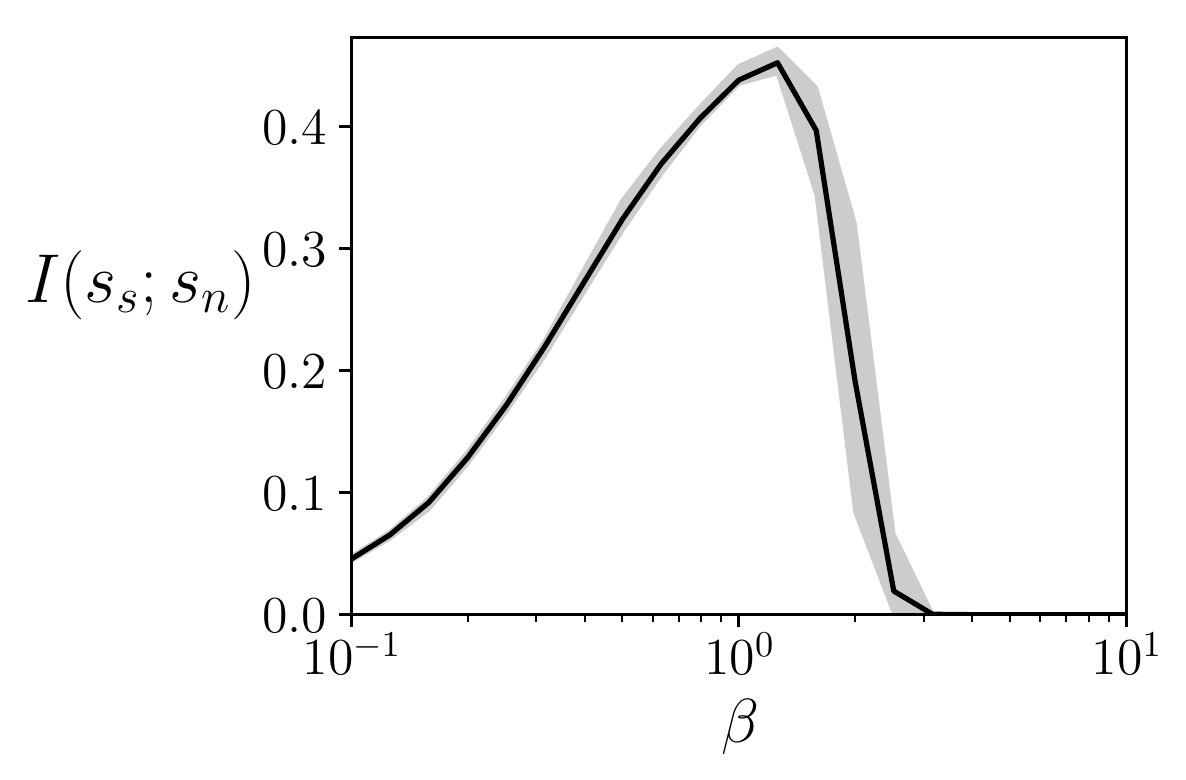}
 \end{tabular}
\end{center}
\caption{Signatures of criticality for 10 different agents in the Mountain Car (left) and Acrobot (right) embodiments. \textbf{(A-B)} Ranked probability distribution function of the Ising models.  The real distribution is compared with a distribution following Zipf's law, (i.e. $P(s) = 1/rank$, dashed line). We observe a good agreement between the model and Zipf's law, suggesting critical scaling. \textbf{(C-D)} Entropy of the neural system $H(s)$. A transition is observed near the operating temperature. \textbf{(E-F)} Mutual information between sensors and neurons $I(s_s;s_n)$. In all figures the mean of all agents is shown as a solid line, while the region between maximum and minimum values showed by the different agents is shown as a grey area.}
\label{fig:criticality}
\end{figure}

First, in order to test whether the agents are being driven near a critical point, we analyze signatures of critical behaviour in the neural controller of the agent. 
Counting the occurrence of each possible state of all the neurons of the agents (including sensor, hidden and motor neurons) we can compute the probability distribution of the Ising model $P(s)$.
We observe that all agents approximately follow a Zipf's law at $\beta=1$ for the Mountain Car (Figure \ref{fig:criticality}.A) and Acrobot embodiments (Figure \ref{fig:criticality}.B), with error bars in a very narrow range.
The power-law distribution of neural activation patterns suggests that the neural controller of the agents is operating near a critical point.

Since the neural network is now connected to an environment that is driven by deterministic mechanics, the probability distribution of the Ising neural controller is no longer described by equation \ref{eq:Ising}. Thus, classical indicators of criticality as the divergence of the heat capacity are not directly calculable from the energy of the model.
However, we can look for other indicators related to second order phase transitions, computed from the probability distribution $P(s)$ calculated from simulation. One indicator can be the behaviour of order parameters such as the entropy of the system, a transition around the critical temperature.
For example, if we compute the entropy of the probability function of the neural controller $H(s) = -\sum_x P(x) \log{P(x)}$ for different values of $\beta$  we observe that the agent is near an order-disorder transition (Figure \ref{fig:criticality}.C,D) (as well, some Mountain Car agents present a smaller transition at larger values of $\beta$, showing that interesting behaviours can also arise in the ordered phase).

In lattice Ising models, critical points are also indicated by other measures such as a peak in the mutual information found in the system  \citep{barnett_information_2013}. Mutual information is defined as:
\begin{equation}
     I(x,y) = \sum_{x,y} P(x,y) \log \frac{P(x,y)}{P(x) P(y)}
\end{equation}
In our case, we measure the mutual information between the set of sensor units $s_s$ and the set containing the rest of the neurons of the controller $s_n$.
The objective is to capture how much information of the sensorimotor coordination of the robot is captured by the system as a whole instead of being contained in the variables alone. In Figure \ref{fig:criticality}.E,F we can observe how mutual information between sensors and neurons has a peak very close to $\beta=1$, suggesting that agents are poised in a point of the parameter space maximizing the exchange of information between the agent and its environment.

These results suggest that the agent's neural controller is operating near a point of criticality, resembling a second order phase transition. As we will see now, no only the agent's neural controller presents indicators of critical activity, but also the behaviour of the agent as a whole.

\begin{figure}[ht]
\begin{center}
 \begin{tabular}{lll}
  \textbf{A} & \textbf{B} & \textbf{C} \\
 \includegraphics[width=2.5cm]{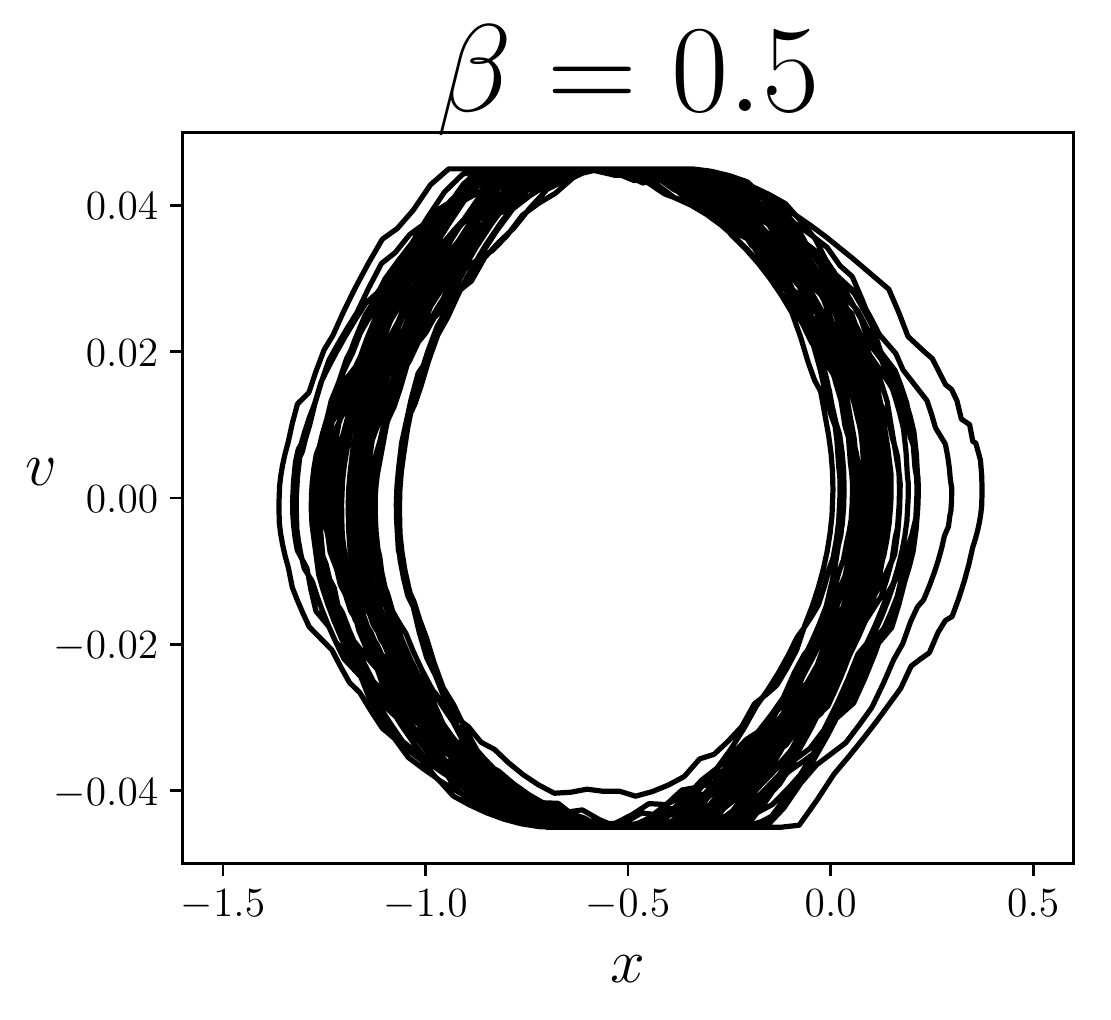} &
 \includegraphics[width=2.5cm]{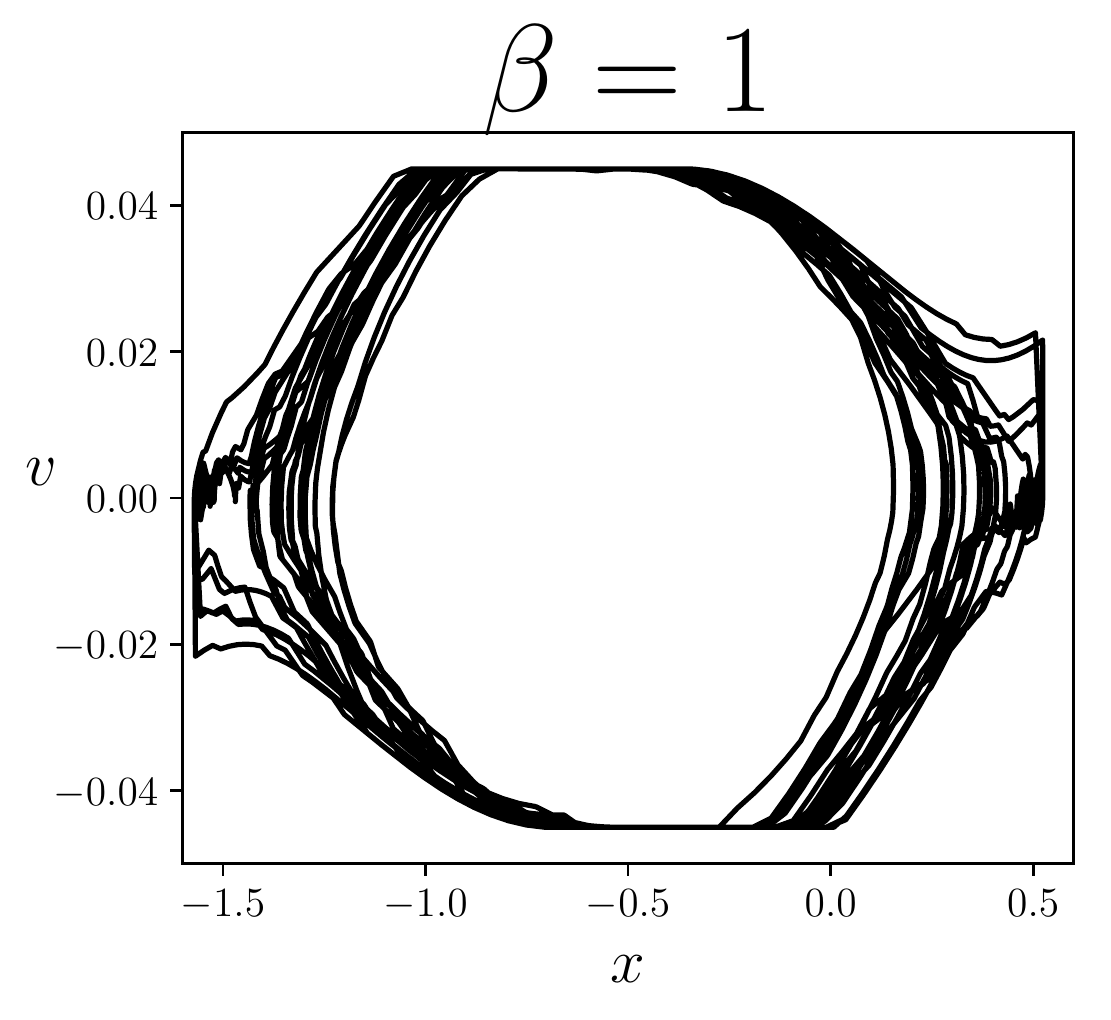} &
 \includegraphics[width=2.5cm]{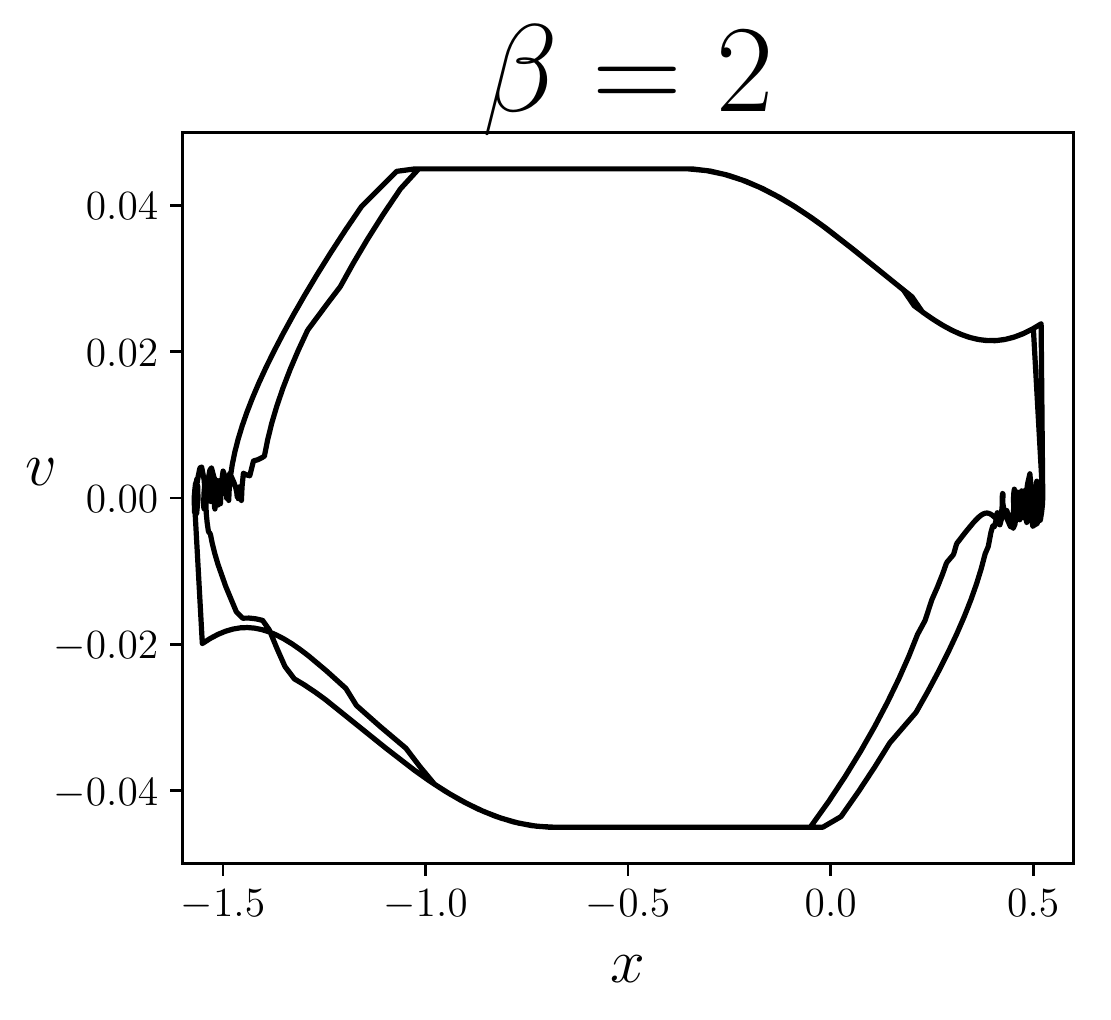} \\
 \includegraphics[width=2.5cm]{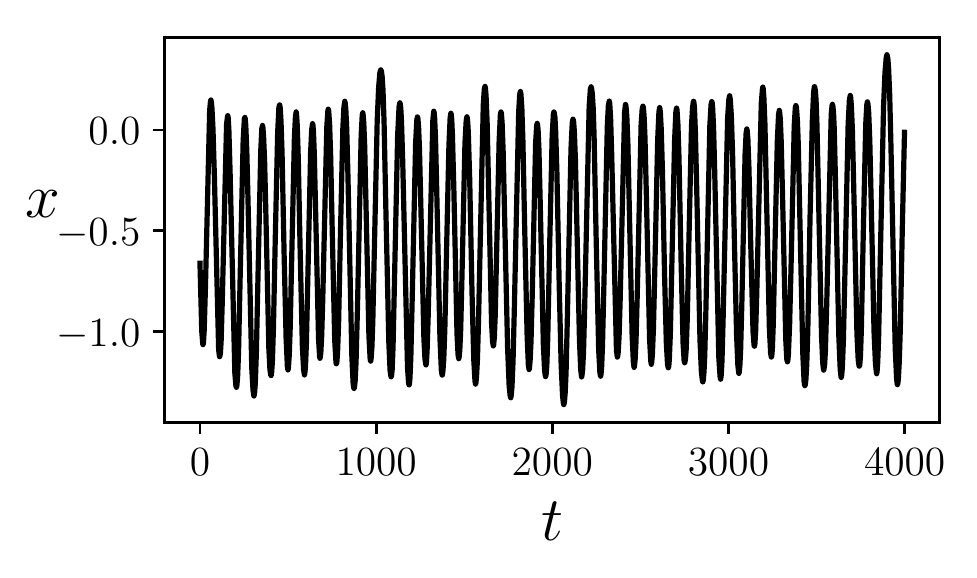} &
 \includegraphics[width=2.5cm]{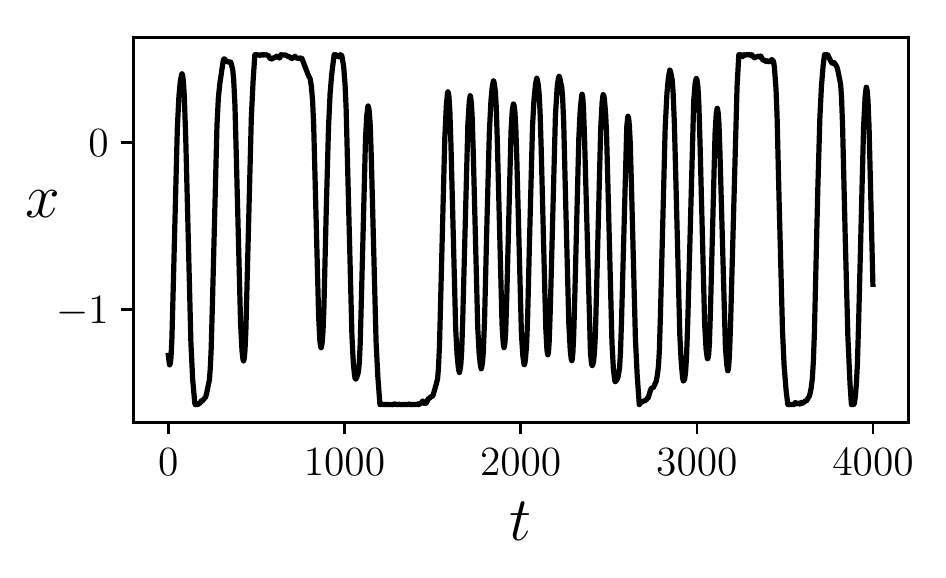} &
 \includegraphics[width=2.5cm]{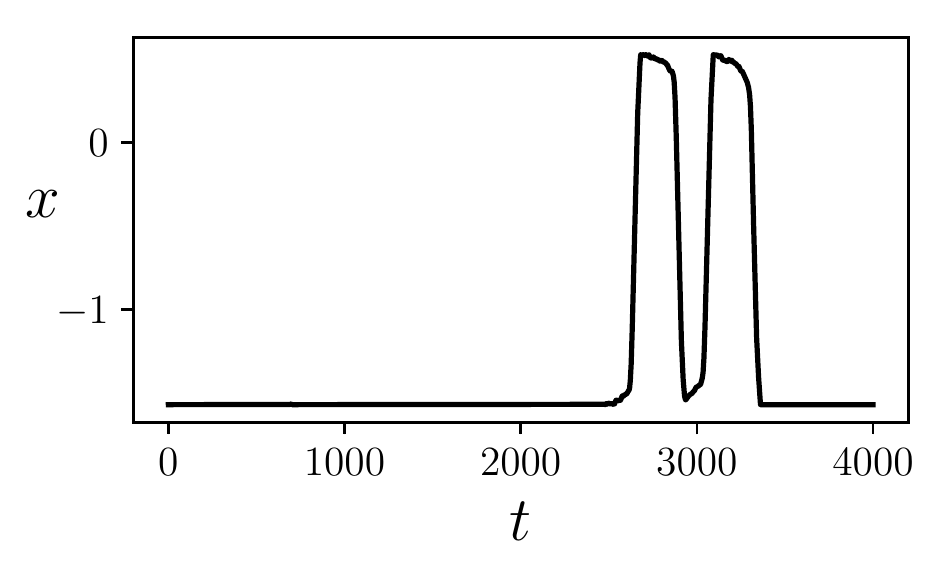} \\
 \end{tabular}
\end{center}

\begin{center}
 \begin{tabular}{lll}
  \textbf{D} & \textbf{E} & \textbf{F} \\
 \includegraphics[width=2.5cm]{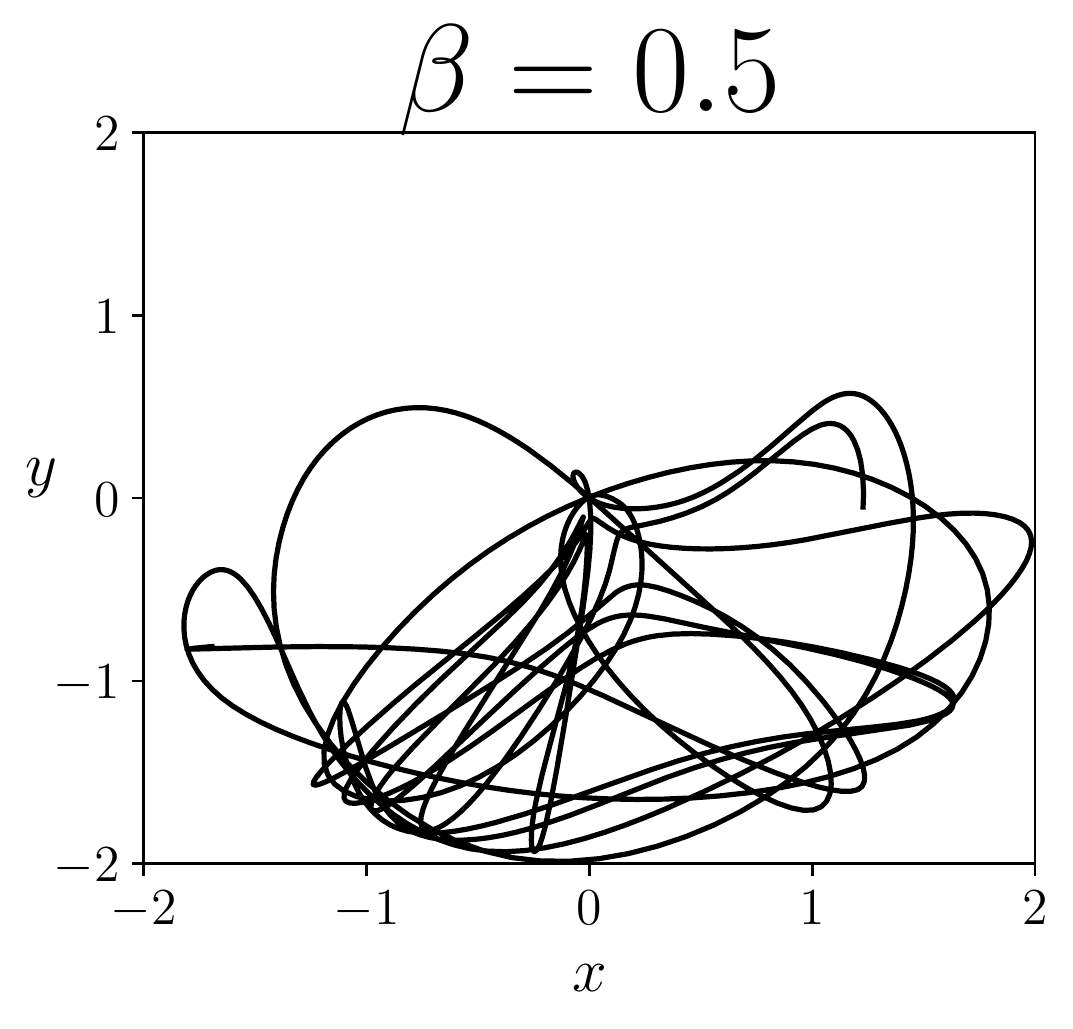} &
 \includegraphics[width=2.5cm]{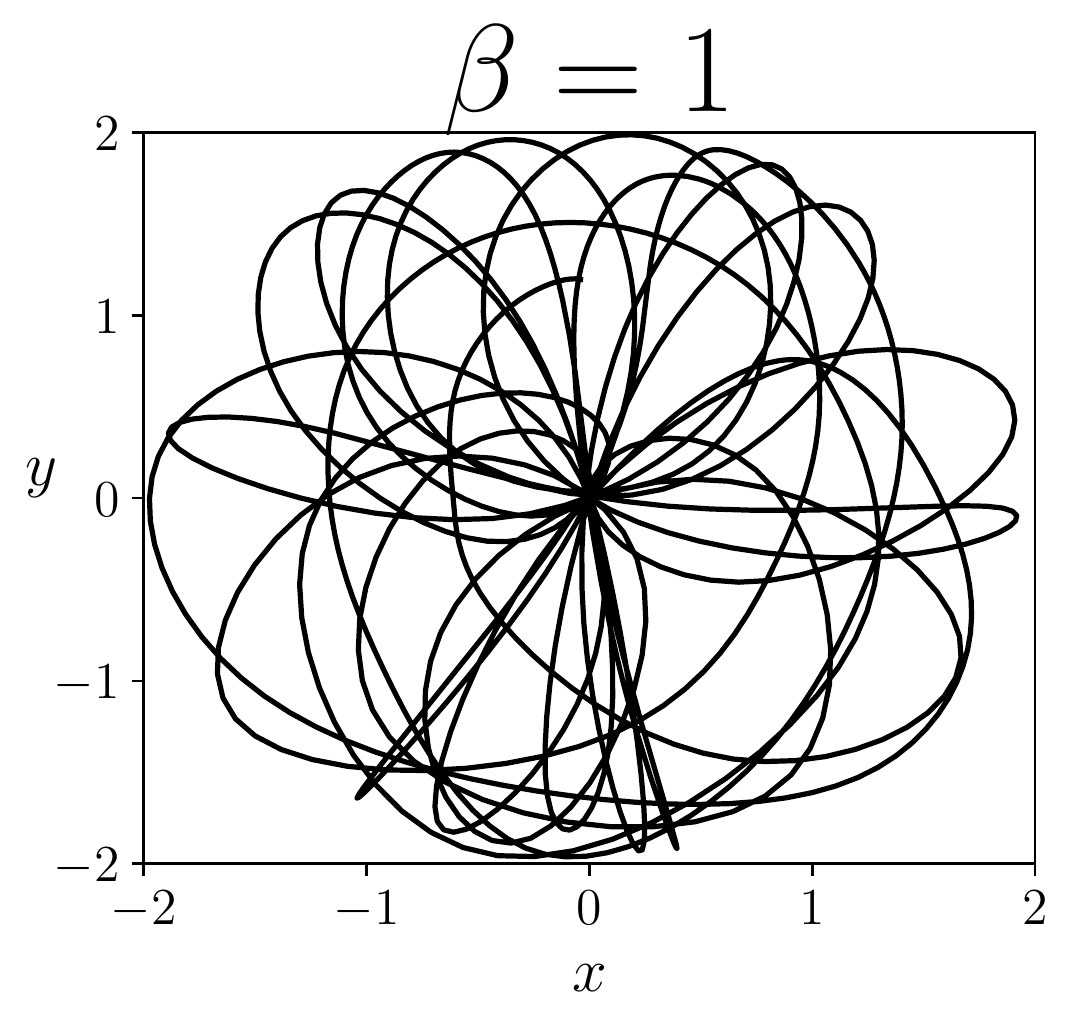} &
 \includegraphics[width=2.5cm]{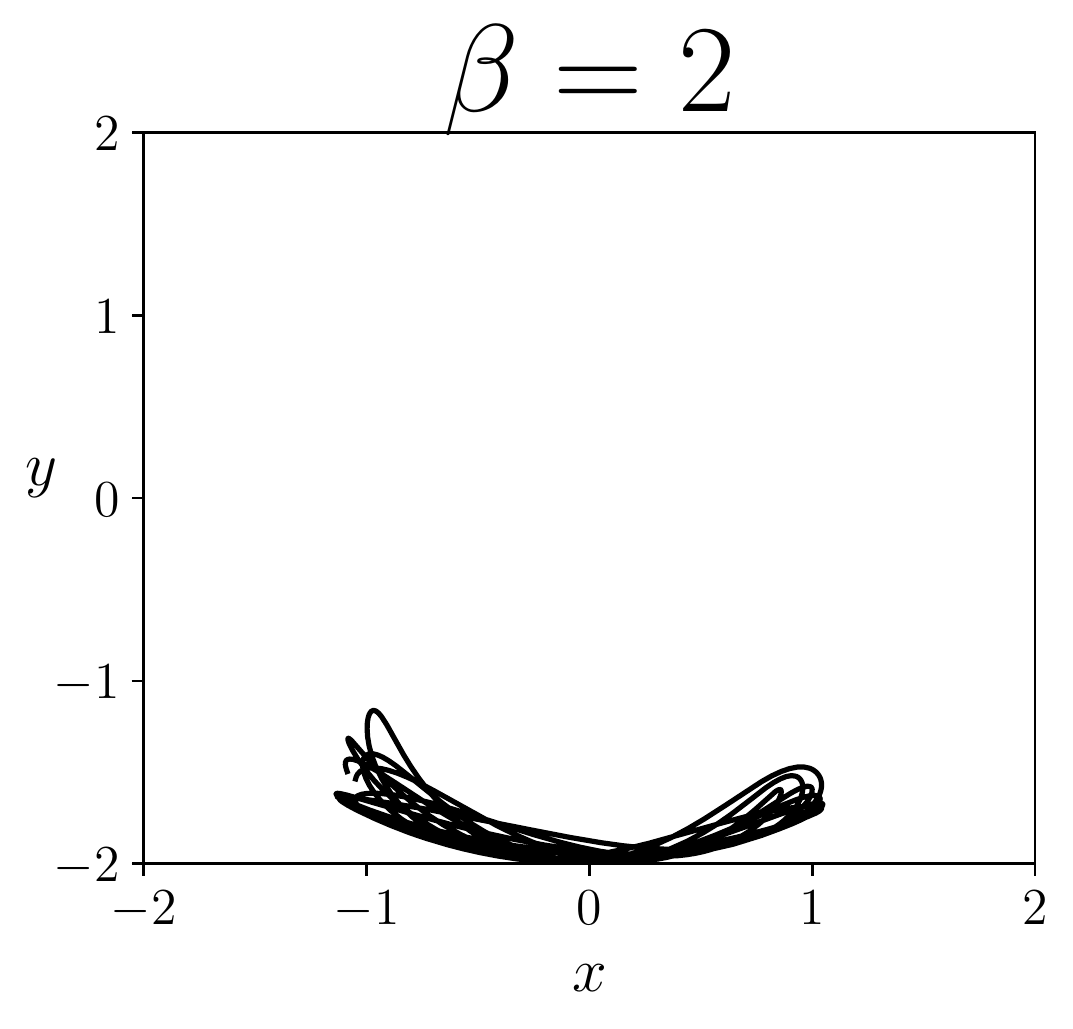} \\
 \includegraphics[width=2.5cm]{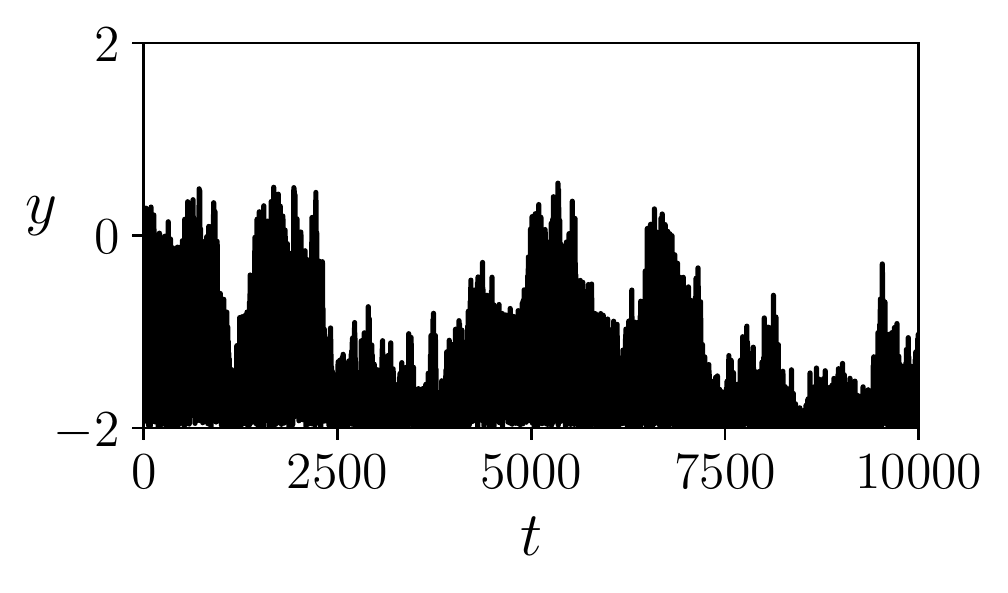} &
 \includegraphics[width=2.5cm]{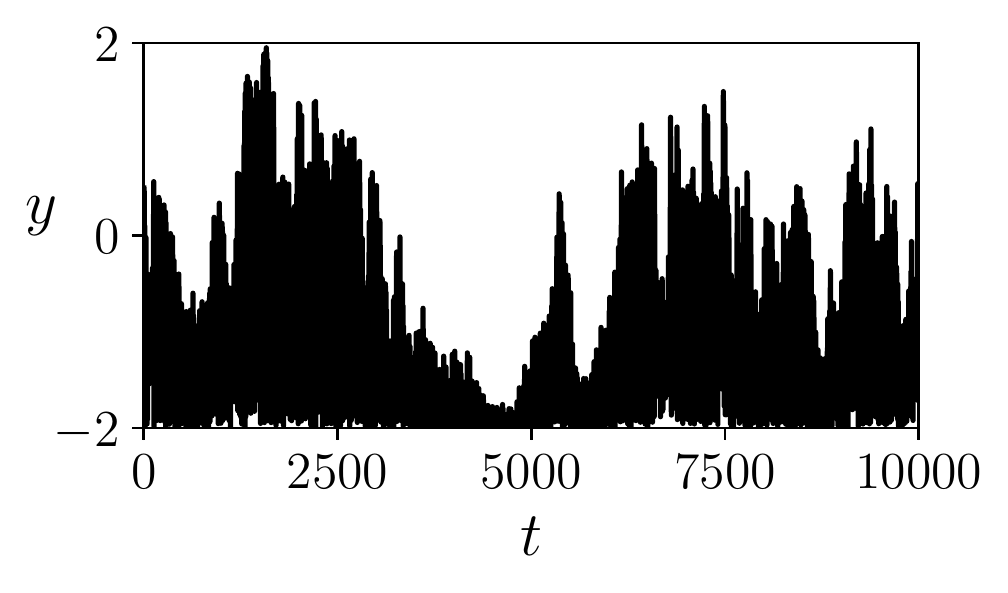} &
 \includegraphics[width=2.5cm]{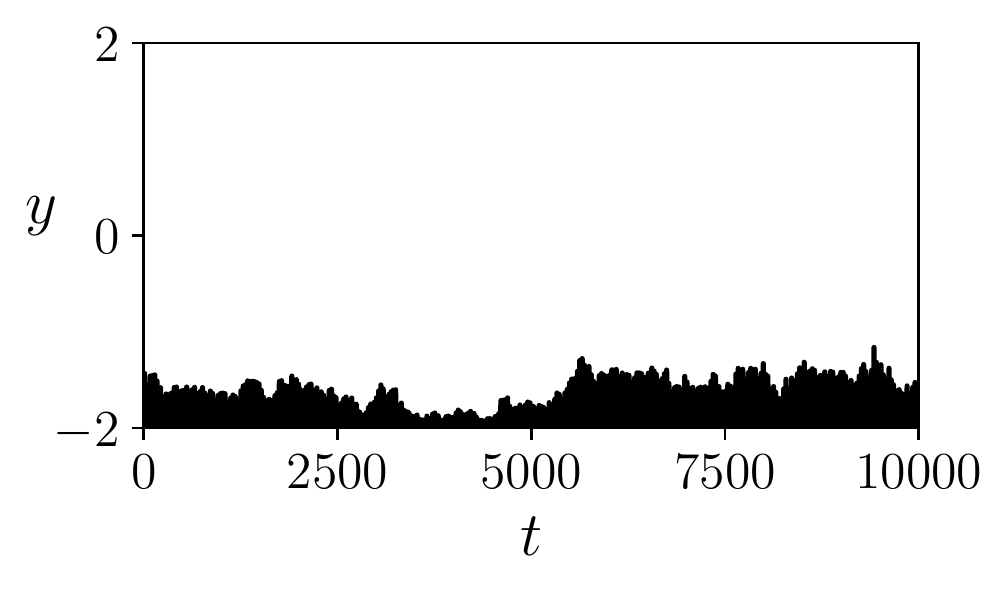} \\
 \end{tabular}
\end{center}
\caption{Transition in the behavioural regime of the agents. We show the behaviour of two agents with different values of $\beta$ for the Mountain Car (\textbf{A}, \textbf{B}, \textbf{C}) and Acrobot (\textbf{D}, \textbf{E}, \textbf{F}) embodiments. We observe that $\beta=1$ is a transition point between two modes of behaviour in both agents.}
\label{fig:transitions}
\end{figure}

\subsection{Behavioural transitions in the parameter space}
What does it imply for the agent to poise its neural controller at a critical point? It should be remarked here that our agents are given no explicit goal but they only tend to behavioural patterns maintaining a distribution of correlations randomly sampled from the distribution in Figure \ref{fig:Correlations}.A. Thus, we  explore the effects of transiting the critical point of the neural controller observing the different behavioural modes of the agent in the parameter space by changing the value of $\beta$. The behaviour of the car can be described just by the position $x$ and speed $v$ at different moments of time. As well, the position of the tip of the Acrobot's links shows a good image of the system's behaviour.

\begin{figure}[ht]
\begin{center}
 \begin{tabular}{ll}
  \textbf{A} & \textbf{B} \\
 \includegraphics[width=3.7cm]{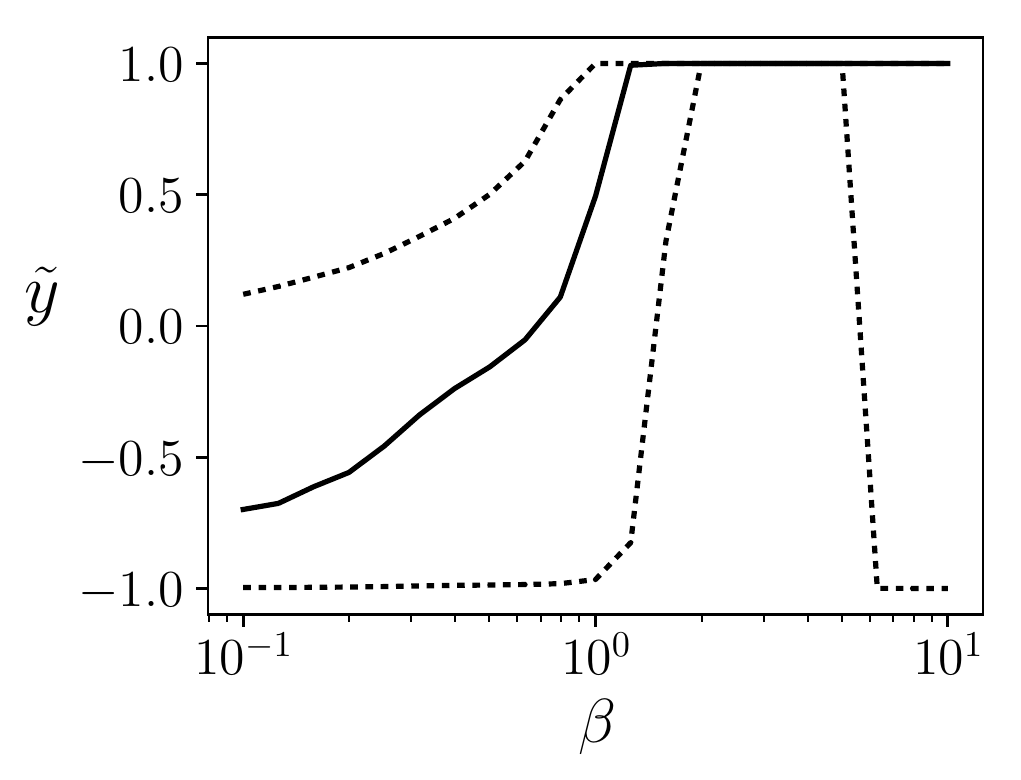} &
 \includegraphics[width=3.7cm]{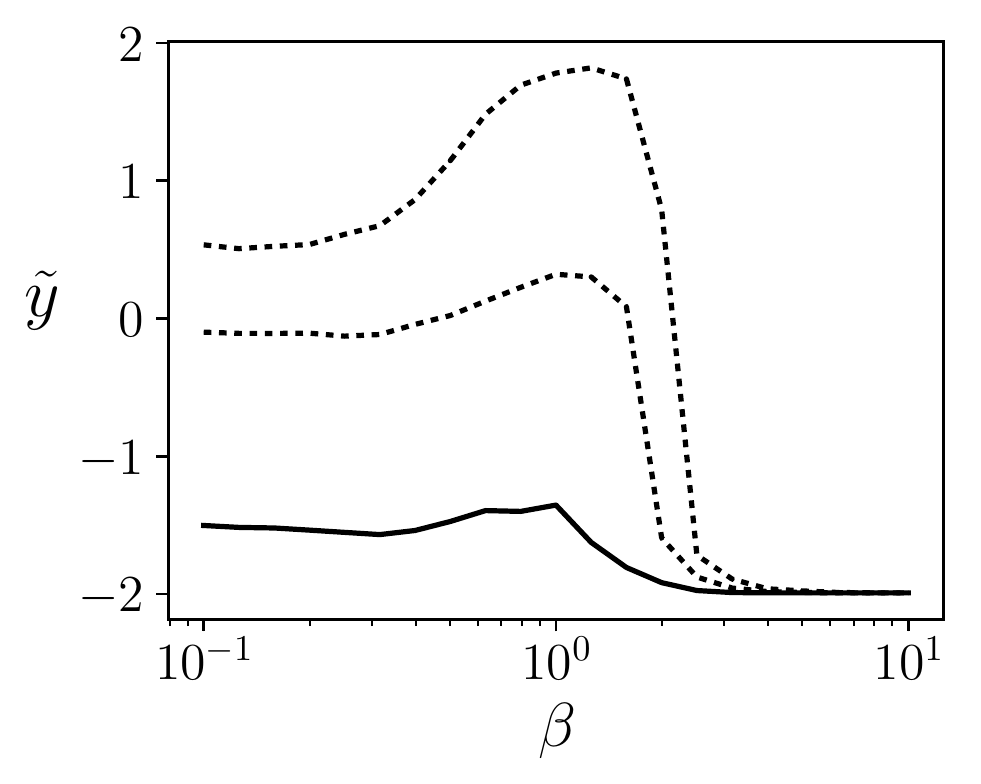}
 \end{tabular}
\end{center}
\caption{Behavioural transitions in the \textbf{(A)} Mountain Car and \textbf{(B)} Acrobot environments, showing the median height $\tilde y$ of the agent, together with the $5\%$ and $95\%$ percentiles in the Mountain Car and the $5\%$ and the $95\%$ and $99.9\%$ percentiles in the Acrobot (dotted lines).} 
\label{fig:behavioural-transition}
\end{figure}

In Figure \ref{fig:transitions}.A-C we can observe the behaviour of the Mountain Car for $\beta=\{0.5,1,2\}$ respectively. We observe that for values of $\beta$ lower than the operating temperature, the agents are not able to reach the top of the mountain. On the other hand, when $\beta$ is higher, the agents present more `rigid' trajectories going form one top of the mountain to the other. At $\beta=1$ the agent is able to reach the top of the mountain (note that the peaks of the mountain are located at $x=-\pi/2$ and $x=\pi/6$) while displaying larger behavioural diversity.
Similarly, in Figure \ref{fig:transitions}.D-F, we observe that the Acrobot at $\beta=1$ displays a diverse range of behaviours, being able to reach to the top of the plane while, when $\beta$ is lowered or increased, it drifts to other behavioural modes in less diverse regimes.
Furthermore, if we compute the median value of height $\tilde y$ for both agents at different values of $\beta$ (Figures \ref{fig:behavioural-transition}), we observe that there is a transition in the parameter space around $\beta=1$, in which both agents maximize the diversity of positions reached in their environment. This suggests that there is a behavioural transition connected with the phase transition of the neural Ising controller, in which the agent maximizes the dynamic range of inputs of the neural controller.

\section{Discussion}
Recapitulating the main ideas presented so far, we have tested how taking a set of correlations chosen at random from a distribution generated by a lattice Ising model at a critical point, we can construct a new model that is also near a critical point of its parameter space. Moreover, imposing an embodied agent to maintain these correlations using a simple learning rule -- as a sort of organizational homeostasis -- drives the agent to a critical point, which coincides with behavioural transitions of its parameter space. 
This suggests the possibility that criticality could be diagnosed and even induced directly from the maintenance of a given distribution of correlations rather than modelling a precise mechanistic structure. Also, criticality could be caused by quite simple mechanisms only relying on local information, maintaining specific correlations around a given value. Here we have implemented the mechanism as a simple Boltzmann Learning process, but other rules could have the same effect, as the combination of Hebbian and anti-Hebbian tendencies in specific ratios.

In our model, we only require the system to maintain a distribution of relations between the components of the system. This connects with systemic approaches to biology interested not in specific or intrinsic components of biological systems but in the networks of relations and processes \citep{bernard_introduction_1957, rosen_relational_1972, ashby_principles_1962}, and it is also in line with notions of relational invariance as Piaget's approach on functional invariants in cognitive development \citep{piaget_biology_1971} or Maturana and Varela's ideas of autopoietic machines, defined as homeostatic systems that maintain constant its own organization as a network of relations between components \citep{maturana_autopoiesis_1980}.

Assuming a similar systemic perspective, we have derived learning rules for a system that drives itself near a critical point by maintaining an invariant structure of correlations roughly defined by a critical exponent $1/r^\eta$. 
This promotes a different perspective on criticality. 
In our model, the distribution of correlations is not the consequence of criticality in a specific topology, but the cause driving an indeterminate topology to a critical point. 
The question now could be whether imposing connections derived from a $1/r^\eta$ function is a strong assumption or implies particularly exigent circumstances. We do not think so, since power law functions can be naturally generated by simple rules of preferential attachment favouring `rich-get-richer' cumulative inequalities \citep{greenwood_inquiry_1920}, or directly as a natural consequence of certain geometries of space \citep[as e.g. gravitation laws, see][]{barrow_new_2002}.

In this way, our model only assumes that a system is going to adapt to preserve an internal network of relations. It emphasizes the maintenance of organizational structures capable of reproducing living systems behaviors being in opposition with the ones relying on internal models of the external source of sensory input. 
Thus, this contrasts with other approaches which have focused in understanding criticality as a strategy to effectively represent a complex and variable external world, for example discovering criticality in predictive coding or deep learning architectures dealing with complex inputs  \citep{friston_perception_2012, lin_critical_2016, hidalgo_information-based_2014}. In those cases, an internalist view is assumed, where the neural controller is representing structures of the external world, whose complexity may be the cause of criticality being present in the neural controller. 
Instead, our approach is agnostic about what are the inputs or the external world of an organism, and deals only with how an agent rearranges its internal structures facing different environments. 

The agents presented here are not specifically designed for a particular problem. In simple terms, our agents generate (preserving the same internal neural organization) a wide variability and richness of behaviours (avoiding both disorder and explosive and indiscriminate propagation) that permits to explore the space of parameters and eventually to achieve solutions of problems for which it was not designed. The empirical evidence of experiments shown here supports this idea. 
A parallel could be established with the concept of play, which can be understood as a `rule-breaker' activity of the constraints of stable and self-equilibrating regime of behaviours, that is not directly required from the environment and without concrete goals \citep{di_paolo_horizons_2010}. 
A model as the one presented here could be used for exploring life-like autonomous behaviour without the need of explicit internal representations, goals, or rule-based behaviour. 
Conceptual models of critical activity based in the maintenance of a system's relational invariants could help developing a synthetic route towards the exploration of adaptive and embodied criticality. 

\section{Acknowledgements}

Research was supported in part by the Spanish National Programme for Fostering Excellence in Scientific and Technical Research project PSI2014-62092-EXP and by the project TIN2016-80347-R funded by the Spanish Ministry of Economy and Competitiveness.

\footnotesize
\section{Materials and methods}

The source code implementing the learning rule in the different examples is freely available at \url{https://github.com/MiguelAguilera/Criticality-as-It-Could-Be}.

\let\oldthebibliography\thebibliography
\let\endoldthebibliography\endthebibliography
\renewenvironment{thebibliography}[1]{
  \begin{oldthebibliography}{#1}
    \setlength{\itemsep}{0.6em}
    \setlength{\parskip}{0em}
}
{
  \end{oldthebibliography}
}

\footnotesize
\bibliographystyle{apalike}
\bibliography{REFS-ECAL17CriticalHomeostasis} 

\end{document}